# Staged deployment of interactive multi-application HPC workflows

Wouter Klijn*, Sandra Diaz-Pier*, Abigail Morrison*§¶, Alexander Peyser*
\* SimLab Neuroscience, Jülich Supercomputing Centre (JSC), Institute for Advanced Simulation,
Forschungszentrum Jülich GmbH, JARA, 52425 Jülich, Germany
§ Computational and Systems Neuroscience, Theoretical Neuroscience & Institute of Neuroscience and Medicine (INM-6)
Institute for Advanced Simulation, Forschungszentrum Jülich and JARA, 52425 Jülich, Germany
¶ Institute of Cognitive Neuroscience, Faculty of Psychology, Ruhr-University Bochum, Bochum, Germany
w.klijn@fz-juelich.de

*Abstract*— **Running scientific workflows on a supercomputer can be a daunting task for a scientific domain specialist. Workflow management solutions (WMS) are a standard method for reducing the complexity of application deployment on high performance computing (HPC) infrastructure. We introduce the design for a middleware system that extends and combines the functionality from existing solutions in order to create a high-level, staged user-centric operation/deployment model. This design addresses the requirements of several use cases in the life sciences, with a focus on neuroscience. In this manuscript we focus on two use cases: 1) three coupled neuronal simulators (for three different space/time scales) with in-transit visualization and 2) a closed-loop workflow optimized by machine learning, coupling a robot with a neural network simulation. We provide a detailed overview of the application-integrated monitoring in relationship with the HPC job. We present here a novel usage model for large scale interactive multi-application workflows running on HPC systems which aims at reducing the complexity of deployment and execution, thus enabling new science.**

## I. INTRODUCTION

Running scientific workflows on a supercomputer can be a daunting task for a scientific domain specialist: nonstandard hardware, application launch via a batch system with unique output files, system specified locations for logs and configuration files. The way applications are deployed reduces user control, and fragments information about application execution over multiple locations. This complicates the identification and resolution of the root causes of failures. The European Flagship Human Brain Project (HBP) [1, 2] is currently developing a number of multi-application co-deployed workflows which are even more challenging due to the additional requirements for interactive steering and deployment on hybrid hardware. The probability of a fault condition grows as a multiplicative function of the number of applications in the workflow and their failure probabilities as in (1) where $P_i$ is component i's failure probability, further exacerbating the challenges for the user.

$$P(system\ failure) = 1 - \prod_i (1 - P_i) \qquad (1)$$

Workflow management solutions (WMS) are a standard method for reducing the complexity of application deployment on high performance computing (HPC) infrastructure. These frameworks allow specification of the work using a formal representation; the framework then takes care of resource allocation and job deployment. Interactivity and improved fault tolerance have been identified as important future research [3], and methods for collecting real-time information on the application's state as well as how to report this back to the end user are subjects of current research [4]. Log files are an additional information source, providing insight into running applications. Collecting, transporting and storing these files is an established practice and integrated frameworks are available (e.g. Elasticsearch, Logstash, Kibana stack (ELK) [5] and Grafana [6]). Further information sources are node and hardware monitoring services like Paraview Gridmonitor [7]. It is not feasible for the intended end user of the HBP workflows, the domain scientist, to debug these systems manually with the information presented to her in such a distributed manner

In this article, we introduce the design for a middleware system that collects information from all these streams at a single location. This centralization will enable a novel end user-centric operation/deployment model for the execution of steered co-deployed multi-application workflows on HPC hardware. Inspired by the checklists that pilots complete before take-off, we introduce the concept of an automated staged workflow deployment procedure. Before and during deployment on HPC hardware, both the system and application states are validated, enabling deployment to be halted when anomalies are detected, thus saving time and valuable computing resources. The end user will have insight into the workflow during deployment and execution by means of dedicated dashboards. The middleware will integrate with and add functionality to existing WMSs but will not be necessary for workflow execution.

The staged deployment procedure needs detailed insight into the real-time state of applications and computing systems. To this end, we propose to greatly extend the functionality, of what we will call in the following the "application companion" (AC). Launched together with the science application, the AC has access to the execution command, system environment and process identifier (PID). This grants the AC access to the application's initial context, thus simplifying error checking. An example of this kind of application is the Pegasus-Kickstart application in the Pegasus framework [8]. Additionally, we propose an invasive but novel information source formed by tracing points located in key locations in the monitored application itself. These three streams provide information at discrete time points. This allows reuse of existing tooling and functionality from software such as ELK and Grafana. A high-level overview of the dashboards and information streams in our middle ware can be found in Fig 1.

## II. RELATED RESEARCH AND SOFTWARE

UNICORE [9] is a framework which provides a HPC-centric workflow definition and cluster scheduling functionality. It includes powerful features such as loops and divergent workflow paths with dependencies on partial results. It performs brokering for HPC resources, deployment of the jobs, and secure communication between processing steps. It also provides access to data stores, and performs file transfers between HPC and the local user as well enabling basic job monitoring. Jobs are typically started serially and interaction with the workflow is limited to starting and stopping.

Pegasus [10] is a workflow management system used in a variety of science domains such as astronomy, bio-informatics and earth sciences [3]. Pegasus supports the execution of workflows on commodity clusters, the cloud and via HPC clustering schedulers. One area of particular focus is resource optimization. Detailed information of the application's execution is collected using Pegasus-Kickstart, a small wrapper that launches applications and collects information about the resource usage and execution state. Efforts are underway to create a version that provides real-time information [11].

Real-time monitoring of running applications is an established practice in cloud computing. Two commonly used frameworks are ELK and Grafana [5, 6]. The primary data source for monitoring are the application log files. Individual log lines are parsed, stored in a (time series) database and subsequently displayed in monitoring dashboards. These methods are seeing a gradual uptake in the HPC setting especially in the context of parallel I/O monitoring [12].

TABLE I. COMPARISON OF KEY FUNCTIONALITY FOR RELATED TOOLS AND FRAMEWORKS

| Tool or framework | Key functionality or requirement | | | |
|---|---|---|---|---|
| | *Workflow manager agnostic* | *Use case dedicated dashboard* | *HPC dedicated* | *Application Internal state monitoring* |
| Pegasus-Kickstart | No | Yes | Yes | No |
| ELK | Yes | Partially | No | Partially |
| Slurm | No | No | Yes | No |
| Proposed middleware | Yes | Yes | Yes | Yes |

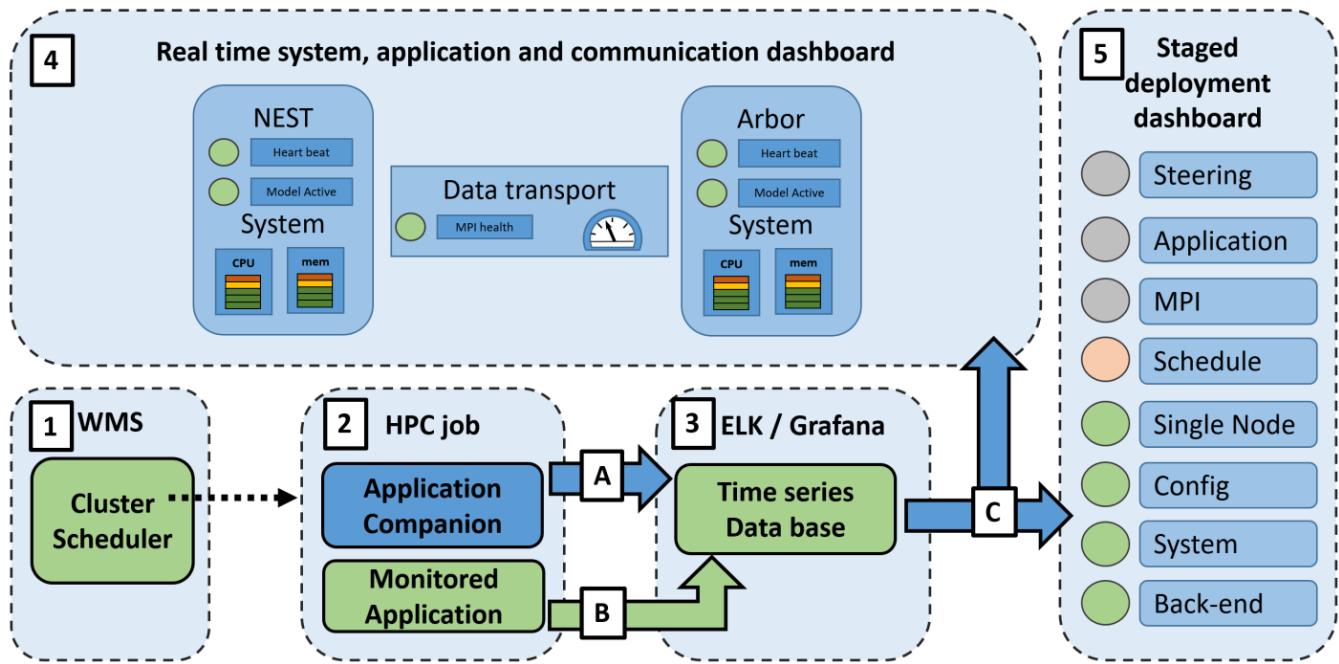

Figure 1. High-level overview of the staged user-centric operation/deployment model and supporting middleware. The WMS (1) launches the Application Companion and application as a regular HPC job (2). Available system and application information (A-B) sources are collected in a single time series database part of ELK or Grafana (3). The database can now be queried and the comprehensive information (C) displayed in dedicated end user dashboards (4,5). A: Resource usage & monitoring plug-ins data collected via the Application Companion. B: Information from application log files. C: Systems and hardware monitoring information. 4: Real time dashboard 5: staged deployment dashboard. Dark blue boxes and arrows in this diagram will be newly implemented, green boxes and arrows are existing functionality and tools which will be integrated.

The software systems and frameworks as discussed in this section are state of the art and we are not proposing to replace or add functionality to the applications itself. The systems are an integral part of our design as can been seen in Fig 1. The added value of our work is to combine functionality and merge information streams. This will allow the end user to be presented with a single end point for monitoring and enable novel operations/deployment models. Table 1 provides a comparison of the proposed middleware system with the related software tools discussed in this section on key functionality and requirements.

### III. USE CASES

The life sciences, and neuroscience in particular, face complex, multiscale problems [13, 14]. The models and interactions bridge many temporal and spatial orders of magnitude and exhibit many degrees of freedom. In the HBP, there is now a push to combine and connect simulators, data sources, and analysis and visualization tools at these different scales. This need to combine heterogeneous software components gives rise to complex workflows of interactive co-deployed applications [15] running on HPC and future hardware systems like neuromorphic computing platforms [16, 17].

We present here two generalized use cases. The first use case integrates multiple neuroscientific simulations at different scales with interactive analysis and visualization and is a prototypical example of a big data processing workflow where intermediate products cannot be stored. The second use-case is a streaming workflow with real world interactions: robot (simulations) and a neuronal network simulator in a closed loop where the network is being optimized with machine learning. These are abstracted versions of workflows currently implemented in the HBP that will need to be production ready for end users in the coming years. This time frame illustrates the current pressing need to develop HPC middleware to support scientific modeling of (multiscale) simulations, and analysis of complex biological systems on HPC [1, 18, 19, 20].

The first use case, illustrated in Fig 2 (top), involves the co-simulation of three simulators: Arbor [21, 22], NEST [23], and The Virtual Brain (TVB) [24], which could be used to solve local field potential calculations at scale, for example as an extension to LFPy [25]. Each simulator has a different characteristic space, time and complexity scale, ranging from morphologically detailed neurons to point neuron models and neural mass models. The models are of such size and complexity that disk IO prevents saving of all output data. The output of the simulators will be sent to the analytical tool Elephant [26, 27]. This tool is currently being extended to allow the creation of summary information without the need for disk access. Finally, by integrating the HBP in-situ pipeline [28], the results of the analysis will be visualized on the graphical front end. This use case has the following set of requirements:

- R1: Robust deployment of several simulation and analysis engines at the same time
- R2: Ability to deploy simulations on HPC infrastructure
- R3: Ability to monitor elements in the workflow and their interactions
- R4: Reproducibility of the experiments.
- R5: Debugging capabilities
- R6: Interactive steering and visualization of each active element in the workflow
- R7: Efficient usage of computational resources
- R8: Easy to understand overview of the status of the workflow at any moment during its execution
- R9: User centric design

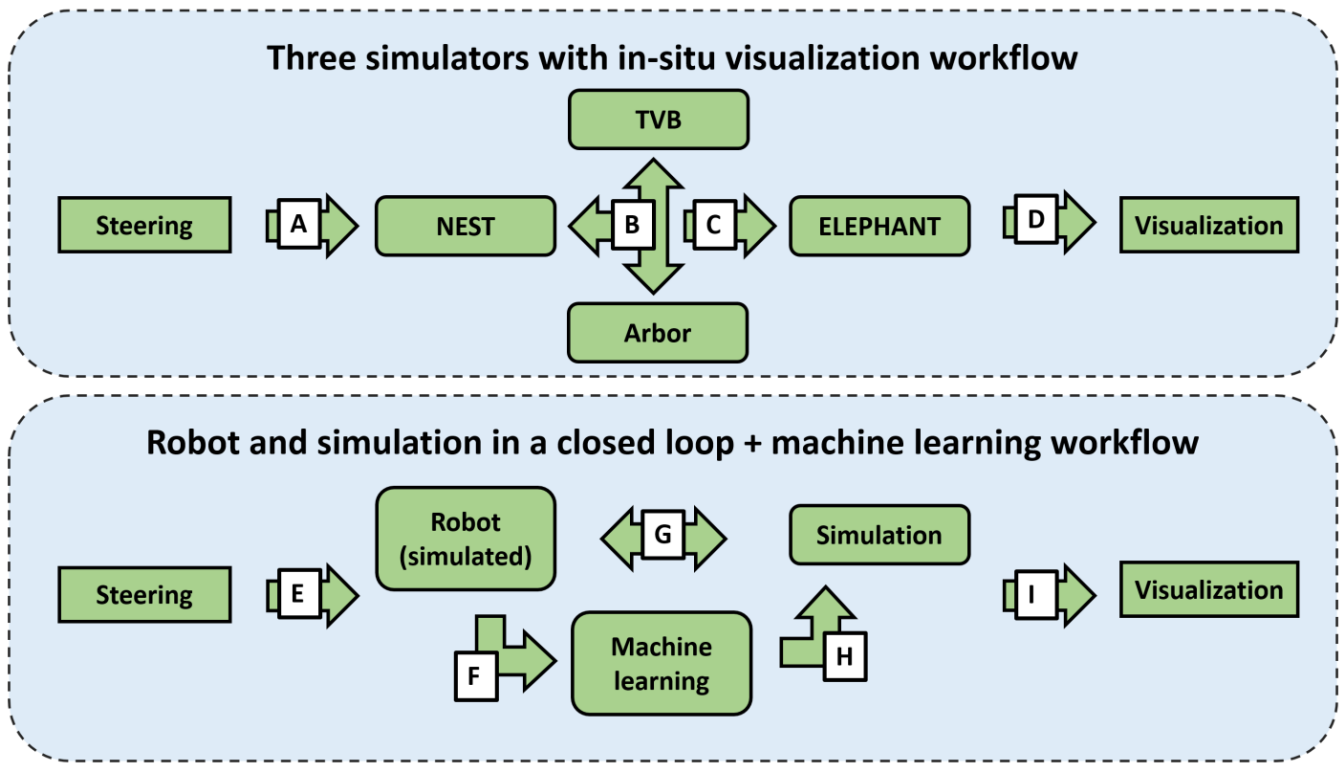

Figure 2. Top: Three simulators with in-situ visualization workflow. The three simulators at different levels of detail, NEST, Arbor and TVB are connected via a dedicated high bandwidth inter-application I/O (B). Other inter-application I/Os are steering commands (A), raw simulation output (C) and visualization I/O (D). Bottom: Robot and simulation in a closed loop plus machine learning workflow. A single simulator, NEST, Arbor or TVB is connected two way to a robot via time critical inter-application I/O (G). Other inter-application I/Os are Sensory and task performance to the machine learning applications (F). The updated network parameters (H). Steering commands to the different applications (E) and task data to be visualization (H). Each individual inter-application I/O and application is a potential source of failure and will be monitored explicitly in our middleware, none of the components or communication channels depicted are implemented as part of our framework.

The second use case introduces interactions with the real world in combination with online machine learning. Machine learning plays an important role in life sciences, e.g. in the broad field of bioinformatics [29]. Our use case is situated in the field of neurorobotics, which focuses on autonomous bio-inspired neural systems, embodied in real or simulated robots. The envisioned (simulated) setup is a mobile robot equipped with an omnidirectional camera and tactile sensors. This robot learns to move through a cluttered environment minimizing the path length without touching obstacles. Data exchange and timing coordination for neuroscience under this real time context, is the goal of the MUSIC [30] interface. This workflow is illustrated in Fig 2 (bottom). Future applications of this workflow include the study of embodied intelligence in robots and brain-machine interfacing in live animals. In addition to the requirements introduced by the first use case, this use case needs:

- R10: Real time interactions
- R11: Ability to keep track of changes in the simulation and analysis engines

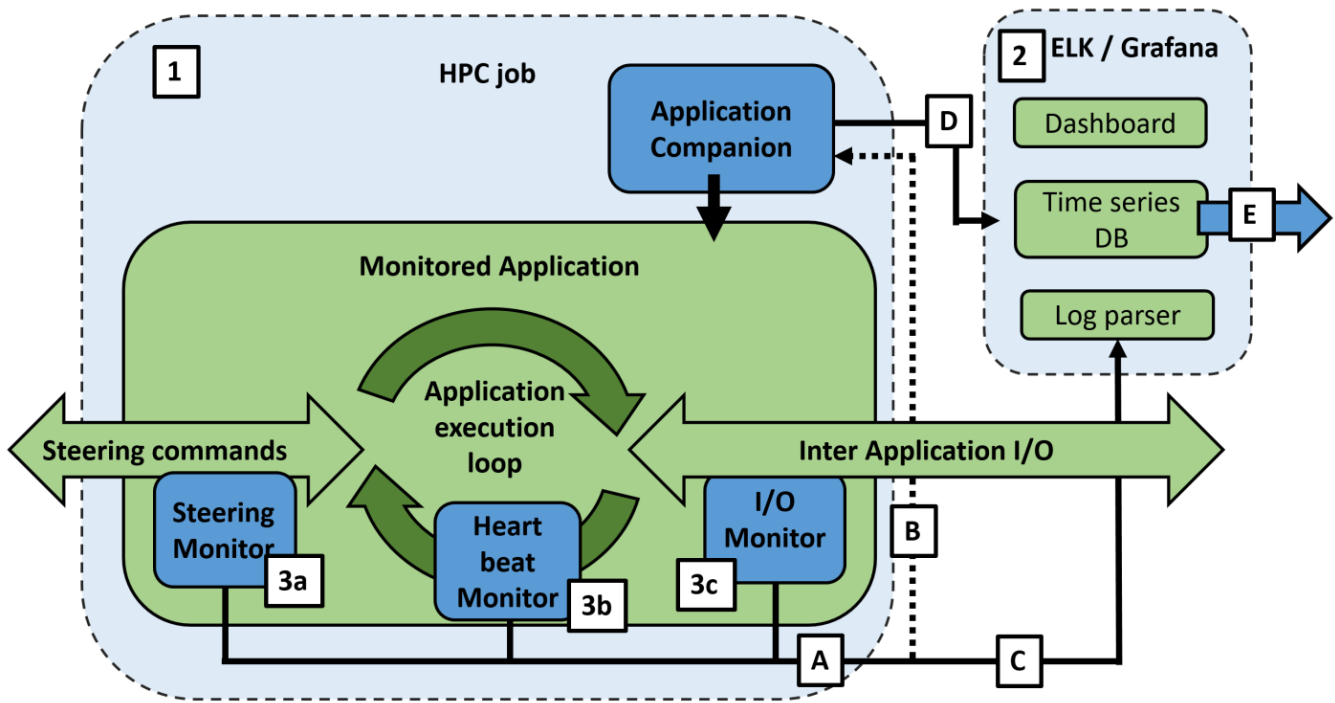

Figure 3. Detailed overview of the application integrated monitoring in relationship with the HPC job (1). Key points in the application (3A-B) are monitored. Collected information (A) will in the first implementation be routed via log files. A later optimization will see this sent (B) via the Application Companion. Application resource usage and execution state will be sent (D) directly to the time series data base, part of the ELK / Grafana stack (2). All collected information can now be queried by dedicated dashboards (E). Key monitoring points are 3a, the Steering Monitor checking the steering commands as send to the application; 3b, the Heartbeat Monitor of ongoing processing in the application. For non iterative application key moments in the processing chain could be monitored; and 3c, the I/O Monitor for big data or time critical application specific communication. Dark blue boxes and arrows in this diagram will be newly implemented, green boxes and arrows are existing functionality or tools which will be integrated.

## IV. DESIGN

### A. Usage model

Conceptually, our usage model isolates the execution of individual applications and communication channels into discrete stages, akin to the OSI standard underlying the protocol stack for internet communications [31]. As a result, it will be possible to do dedicated testing and monitoring of dependencies for each application at specific stages of execution (R1, R2, R3). This facilitates the identification of critical errors and faults at different levels of abstraction [32] (R5). One key element in our deployment model is the addition of a first execution stage, performed on a single node. This entails the implementation of reduced test versions of the workflows to be deployed; limited in size to a single node, but containing all components of the production workflow. This stage verifies dependencies like software libraries, input/output paths, execution parameters, privileges and correct job configuration. Performing these verifications before large-scale HPC deployment results in enhanced robustness, increased success rate and minimizes loss of valuable computational resources (R1, R7). Having successfully obtained a green light from the single node verification step, the HPC execution stage starts, validating multi-node dependencies. A dedicated monitoring console will give the user a comprehensive view of this deployment process (R1, R8, R9). A mock-up of a staged deployment monitor can be found as part of Fig 1.

An additional console will provide operational system health information during the final target stage: a running interactive multi-application science workflow running on HPC. The initial version of the dashboards will be implemented in Grafana or Kibana (the K in the ELK stack). Implementation of a dedicated monitoring application, one that removes as much distracting information from the end user as possible, is deferred to a later implementation stage. Dedicated usability engineering will determine what the most effective representation is. A mockup of a dedicated dashboard can be found in Fig 1. A usability driven design will assure the appropriate level of detail and information content for the end user (R8, R9).

### B. Collecting real time system and application state

The usage model, as detailed in the previous sections, depends critically on detailed information of the system and internal application states (R3). This information will be collected from three sources: application log files, system and application execution data as provided by the operating system, and finally application-integrated monitoring. Information collection from log files is the primary goal of both ELK and Grafana. One important component for both systems is the database. This database is typically, a time series based system, optimized for storing and retrieving series of data points in time.

In the following, we discuss the functionality of the application companion (AC). This is an application which will receive the PID of the main application to be launched together with that main HPC application. AC has a role comparable to the Pegasus-Kickstart application [8, 11]. It collects information provided by the OS and has access to the application exit state. Pegasus-Kickstart is tightly integrated with the WMS. Our AC is also executed concurrently with the monitored application. It wraps the application executable in the deployment script. This allows the CA to execute the application as a subprocess. Giving the CA access to operating system (OS) process information based on the PID. This deployment model additionally provide access to the exit code of the application in a non-invasive manner. The companion app will generate data points, in time, which will be offloaded into the time series database of ELK or Grafana stack allowing reuse of existing tooling.

A second role for the application companion is as a relay point for information collected from application integrated monitoring (R8). The two information streams discussed until now, logfiles and operating system information, are collected in a non-invasive manner. It is already generated by the application (and OS) and in our system is collected and presented in a unified manner. In contrast, the third source of system information, application integrated monitoring, necessitates changes in the application code. These changes consist of code blocks to be added at strategic places in the code. They provide direct insight into essential steps in the application. In the first design, we are implementing the following end points: a simulation loop heartbeat, dedicated Message Passing Interface (MPI) / HPC transport monitoring, and monitoring of the steering commands from the interactive front end (R6, R10, R11). The applications in the use cases are all iterative in nature. For non iterative applications key moments in the processing chain could be monitored. Implementation of this functionality is deferred to a later stage.

In the first implementation of the application integrated monitoring collected information will be added to the existing logfile data stream. But this type of monitoring should have minimal impact on the application itself. In later implementation this information will be routed via the application companion. Data exchange would preferably be done with direct memory access or a point to point messaging based protocol. It should not call database access routines or introduce delays into the monitored applications. The interfaces for the code blocks are developed in close collaboration with the application developers. Fig 3 shows a detailed diagram of the AC in relationship to the monitored app including the expected information streams.

The integration of the middleware will be done in such a way as to only add functionality to existing WMS systems. It should in principle, be possible to launch the workflow without the middleware. Workflow definition and resource allocation is handled by the WMS. The companion app is launched concurrently with monitored application where it receives the command line and PID of the launched application. The standard functionality of ELK or Grafana already gives access to the log files. The companion app will implement the interfaces to the time-series database to expose the front end to the additional system and application internal monitoring data. Fig 1 gives a bird eye view of the whole system.

## V. DISCUSSION

We present a novel usage model for large scale interactive multi-application workflows running on HPC systems. The usage model sees a staged deployment of the set of applications and a unified end user dashboard providing insight in the running set of applications. The use case model depends critically on realtime information of both system and application state. To this end we complement the usage model with the design of a simple middle ware. This middle ware is integrated in existing HPC and cloud systems. Building upon existing tooling it enables a novel deployment model for HPC.

The middleware and usage model described in this manuscript can be compared with pilot systems [33]. These are systems that extend functionality of a WMS and are used to deploy workflows on HPC resources. In contrast to the typical pilot system our middleware does not do workflow creation and/or management and it does not function as a resource place holder. The workflows can still be launched without the middleware in place, the middleware functions more like a co-pilot: it relieves the user of the HPC deployed workflow from a burdensome task but leaving the execution of the system (the flying) to the existing WMS.

In this article, we described the technical design and concepts of our middleware. Besides an assessment of the compliance with the requirements of the use cases, explicit validation is only possible in the later development stage. Development tools and practices specifically suited to help in this process are available, e.g. *chaos engineering* which is a devops practice for distributed system management [34]. Chaos engineering employs tools and services to randomly select software components or workflow stages to fail. This reliable source of faults is a valuable tool when testing and validating a fault-monitoring framework. Additionally, tools are available to generate synthetic workloads. Both generic tools [35] and neuroscience application specific tools exist [22, 36]. The combination of these two sets of tools will allow validation of functionality, even in the absence of fully implemented use cases.

A proof of concept is currently being implemented. It is limited to the companion application and minimal integrated application monitoring. Gathered information will be sent to the application log and visualized in the ELK or Grafana graphical front end. The dedicated domain specific monitoring front-ends will be designed in collaboration with the domain scientists and implemented for the production version. Existing close collaboration with the application developers will assure that the application-integrated monitoring will be done in a sustainable manner. The applications in the use-cases are in active development.

One of the main goals of developing such a middleware is to allow modular workflows with HPC requirements to be deployed by a non-IT expert. In order to enable this, our design relies on a technical and social contract developed alongside code interfaces. The middleware described in this manuscript has a large number of external and internal interfaces. These will be designed and matured in close collaboration with the application developers using a software science co-design approach. The development of such contracts allows the integration of future elements into the workflows in a simple yet robust manner and contributes to the standardization of information exchange. The added value of the framework is the combination of existing engineering solutions and methodologies to solve a specific challenge when using HPC: The high barrier of entrance. This design addresses the requirements of a science community with limited experience with HPC and a need for improved usability, robustness, and debugging capabilities of complex interactive workflows to be deployed on a production setting in the near future.

In this design we focus mostly on application logfiles, application monitoring, and the companion app. HPC systems contain more information sources, ParaStation ClusterTools for example collects WMS and Hardware level information. Connections to these external systems will implemented in a future stage. A second future extension of the middleware could be provenance tracking (R4). The monitoring of user steering provides a natural point for collecting and storing the history of interactions between the user and the applications. This would allow future reconstruction of these interactions.

The co-design of hardware and software for scientific purposes is a challenging task. Scientific software is usually not built in such a way that performance and resource usage information can be easily extracted from it. Even more complicated is the case when multiple applications interact. In order to foresee future requirements of scientific workflows, a tool to create estimations of scaling behavior and resource usage is needed. This design enables monitoring of memory consumption, network communications and I/O of complicated workflows at different levels of invasiveness. This usage model can become an important tool in the assessment of the performance of complicated in-silico experiments which otherwise would be very hard to observe.

ACKNOWLEDGMENT

Funded by the Helmholtz Association through the Helmholtz Portfolio Theme "Supercomputing and modeling for the Human Brain". This project has received funding from the European Union's Horizon 2020 research and innovation programme under grant agreement No 720270 (HBP SGA1) and No. 785907 (HBP SGA2).

We want to thank Benjamin Weyers & Wolfram Schenck for many fruitful discussions regarding the topics in this article